\documentclass[%
 reprint,
superscriptaddress,
 amsmath,amssymb,
 aps,
pra,
]{revtex4-2}

\usepackage{graphicx}% Include figure files
\usepackage{dcolumn}% Align table columns on decimal point
\usepackage{bm}% bold math
\usepackage{physics}
\usepackage{amsmath}
\newcommand{\bonnpi}{Physikalisches Institut, University of Bonn, Nussallee 12, 53115 Bonn, Germany}
\newcommand{\geneva}{Department of Quantum Matter Physics, University of Geneva, Quai Ernest-Ansermet 24, 1211 Geneva, Switzerland}

\begin{document}

\title{Breaking strong symmetries in dissipative quantum systems: 
Bosonic atoms coupled to a cavity }

\date{\today}

\begin{abstract}
  In dissipative quantum systems, strong symmetries can lead to the existence of conservation laws and multiple steady states. In this work we investigate a strong symmetry for bosonic atoms coupled to an optical cavity, an experimentally relevant system, generalizing the adiabatic elimination techniques and using numerically exact matrix product state methods. We show that for ideal bosons coupled to the cavity multiple steady states exist and in each symmetry sector a dissipative phase transition occurs at a different critical point. This implies that phases of very different nature can coexist. We find that the introduction of a slight breaking of the strong symmetry by a small interaction term leads to a direct transition from multiple steady states to a unique steady state. We point out the phenomenon of dissipative freezing, the breaking of the conservation law at the level of individual realizations in the presence of the strong symmetry. For a small breaking of the strong symmetry we see that the behavior of the individual trajectories still shows some signs of this dissipative freezing before it fades out for a larger symmetry breaking terms. 
\end{abstract}
\author{Catalin-Mihai Halati}
\affiliation{\bonnpi}
\affiliation{\geneva}
\author{Ameneh Sheikhan}
\affiliation{\bonnpi}
\author{Corinna Kollath}
\affiliation{\bonnpi}
\maketitle

Symmetries play a key role to classifying and unifying the physics occurring in different microscopic systems. A famous example is the universal behavior arising at (quantum) phase transitions. This universal behavior is independent of the microscopic details of the system and can be classified by the symmetries  which are spontaneously broken at the transition. Typically, each symmetry of a Hamiltonian is connected to a conservation law. This has the crucial consequence that the long time state remembers the initial conditions and that the conservation laws need to be considered constructing {\it thermal ensembles} \cite{Balian2007}, the so-called generalized Gibbs ensembles \cite{RigolOlshanii2007}.

Surprisingly, in contrast to the Hamiltonian case, for systems described by the dissipative Lindblad master equations $\pdv{t} \rho = \mathcal{L}(\rho)$, where $\rho$ is the density matrix and $\mathcal{L}$ the Liouvillian, a symmetry of the Liouvillian does {\it not always} imply a conserved quantity and multiple steady states \cite{BucaProsen2012, AlbertJiang2014}. Let the Hermitian operator $\mathcal{O}$ be the generator of the symmetry $\mathcal{U}=\exp(i\phi \mathcal{O})$, with real $\phi$. If the symmetry operator satisfies the condition
$\mathcal{L}\left(\mathcal{U}\rho \mathcal{U}^\dagger\right)=\mathcal{U}\mathcal{L}(\rho)\mathcal{U}^\dagger$,
we have only a so-called weak symmetry. This weak symmetry condition is not sufficient to imply the existence of a conserved quantity or multiple steady states. Only if additionally $\mathcal{O}$ is commuting with both the Hamiltonian and all jump operators $J_m$,
$[\mathcal{O},H]= [\mathcal{O},J_m]=0$,
a so-called strong symmetry exists which implies conservation of $\langle \mathcal{O}\rangle=\tr(\rho O)$ and a multiple steady states. 

Recently, the consequences of the weak and strong symmetries in open systems were discussed in the context of error correction for quantum information theory \cite{LieuGorshkov2020}.

Experimental systems that can be described by a Lindblad master equation are very frequent in the area of quantum optics and solid state systems coupled to light. In many situations, a unique steady state arises. However, in recent years a significant amount of work has been devoted to go beyond this {\it typical} situation and to study the coexistence of several states in such Lindblad systems \cite{KesslerCirac2012, MingantiCiuti2018, Carmichael2015, Weimer2015, BenitoNavarrete2016, SiebererDiehl2013, MunozNori2018, BiondiSchmidt2017, HwangPlenio2018, MendozaArenasJaksch2016, WilmingEisert2017, HannukainenLarson2018, FerreiraRibeiro2019} and the phenomena of bi-/metastability \cite{MacieszczakGarrahan2016, FinkImamoglu2018, CarrWeatherill2013, deMeloWeatherill2016, MendozaArenasJaksch2016,RodriguezBloch2017, LetscherOtt2017, MuppallaKirchmair2018, SchuetzGiedke2013, SchuetzGiedke2014, HrubyEsslinger2018} or intermittency \cite{FitzpatrickHouck2017, BiondiSchmidt2017, HwangPlenio2018, MuppallaKirchmair2018, MendozaArenasJaksch2016}. 
Additionally, steady states with sought-after properties have been constructed employing symmetries of the system, such as steady states with $\eta$-pairing correlations \cite{BernierKollath2013} or state with enhanced currents \cite{ManzanoHurtado2014, LangeRosch2017} or in weakly driven systems \cite{LenarcicRosch2018}.

We show in this work for a realistic experimental system, a cavity coupled to a quantum gas, how the presence of a strong symmetry can lead to the occurrence of multiple dissipative phase transitions in different symmetry sectors.
We identify that the phase transition can occur for different critical values depending on the considered symmetry sector. Thus, for the same physical parameters the nature of the steady state can be very different depending on the initial state of the system. 

We further investigate how in the situation of the slight breaking of the strong symmetry  by an additional term in the Liouvillian the unique steady state is recovered. Thus, the slight breaking causes a drastic response of the system.
We investigate the timescales associated to the process of reaching the unique steady state. Additionally, we show the absence of intermittency, the dissipative freezing \cite{MunozPorras2019}, in the presence of the strong symmetry in single trajectories obtained by the stochastic unravelling of the master equation.  We find that this behavior of the absence of intermittency can approximately survive for an intermediate time when adding a small symmetry breaking term. Whereas for larger symmetry breaking term the different symmetry sectors are no longer a good description of the system.

%\section{Symmetries of the model} 

We consider ultracold bosons confined to a one-dimensional chain coupled to a single cavity mode and transversely pumped with a standing-wave laser beam \cite{HalatiKollath2020}. 
Adiabatically eliminating the excited internal state of the atoms, the dynamics follows the Lindblad equation \cite{CarmichaelBook,BreuerPetruccione2002,RitschEsslinger2013, MaschlerRitsch2008,HalatiKollath2020} 
\begin{align}
\label{eq:Lindblad}
 \pdv{t} \rho &= \mathcal{L}(\rho)=-\frac{i}{\hbar} \left[ H, \rho \right] + \frac{\Gamma}{2}\left(2a\rho a^\dagger-a^\dagger a \rho-\rho a^\dagger a\right), 
\end{align}
where $\mathcal{L}(\rho)$ is the Liouvillian.
The bosonic operators $a$ and $a^\dagger$ are the annihilation and creation operators for the photon mode of the cavity. The dissipator with strength $\Gamma$ represents the losses from the cavity due to the imperfections of the mirrors. The Hamiltonian, $H=H_0+H_{\text{int}}$, is given by \cite{RitschEsslinger2013, MaschlerRitsch2008, NagyDomokos2008}
\begin{align} 
\label{eq:Hamiltonian}
 H_0=& \hbar\delta a^\dagger a -\hbar\Omega ( a + a^\dagger) \sum_{j} b_{k_j}^\dagger b_{k_j+\pi~(\mathrm{mod}~2\pi)}\\
 &-2J \sum_{j} \cos(k_j) b_{k_j}^\dagger b_{k_j}\nonumber\\
H_{\text{int}}=&\frac{U}{2} \sum_{l} n_{l}(n_{l}-1). \nonumber
\end{align}
The cavity mode is described by the first term in $H_0$, in the rotating frame of the pump beam, where $\delta=\omega_c-\omega_p$ is the detuning between the cavity mode and the transverse pump beam. The operators $b_{k_j}$ and $b_{k_j}^\dagger$ are the bosonic annihilation and creation operators of the atoms with the unitless momentum $k_j=\frac{2 \pi j}{L}$ and $j=1,\dotsc,L$, assuming periodic boundary conditions. In the numerical results we considered open boundary conditions \cite{sup}. $J$ is the tunneling amplitude of the atoms and $U\geq 0$ the strength of the on-site interaction, where $l$ denotes the site of the chain and $n_l$ the atomic density.
$L$ denotes the number of sites of the bosonic chain and the total number of bosons is $N$.  The coupling between the atoms and the cavity field introduces a change of the momentum $k_j$ and $k_j+\pi~(\mathrm{mod}~2\pi)$. This is due to the periodicity of cavity mode which has twice the periodicity of the lattice spacing within the chain. The commensurability of the chain and the cavity field is an important condition for the realization of the strong symmetry.

\begin{figure}[hbtp]
\centering
\includegraphics[width=.48\textwidth]{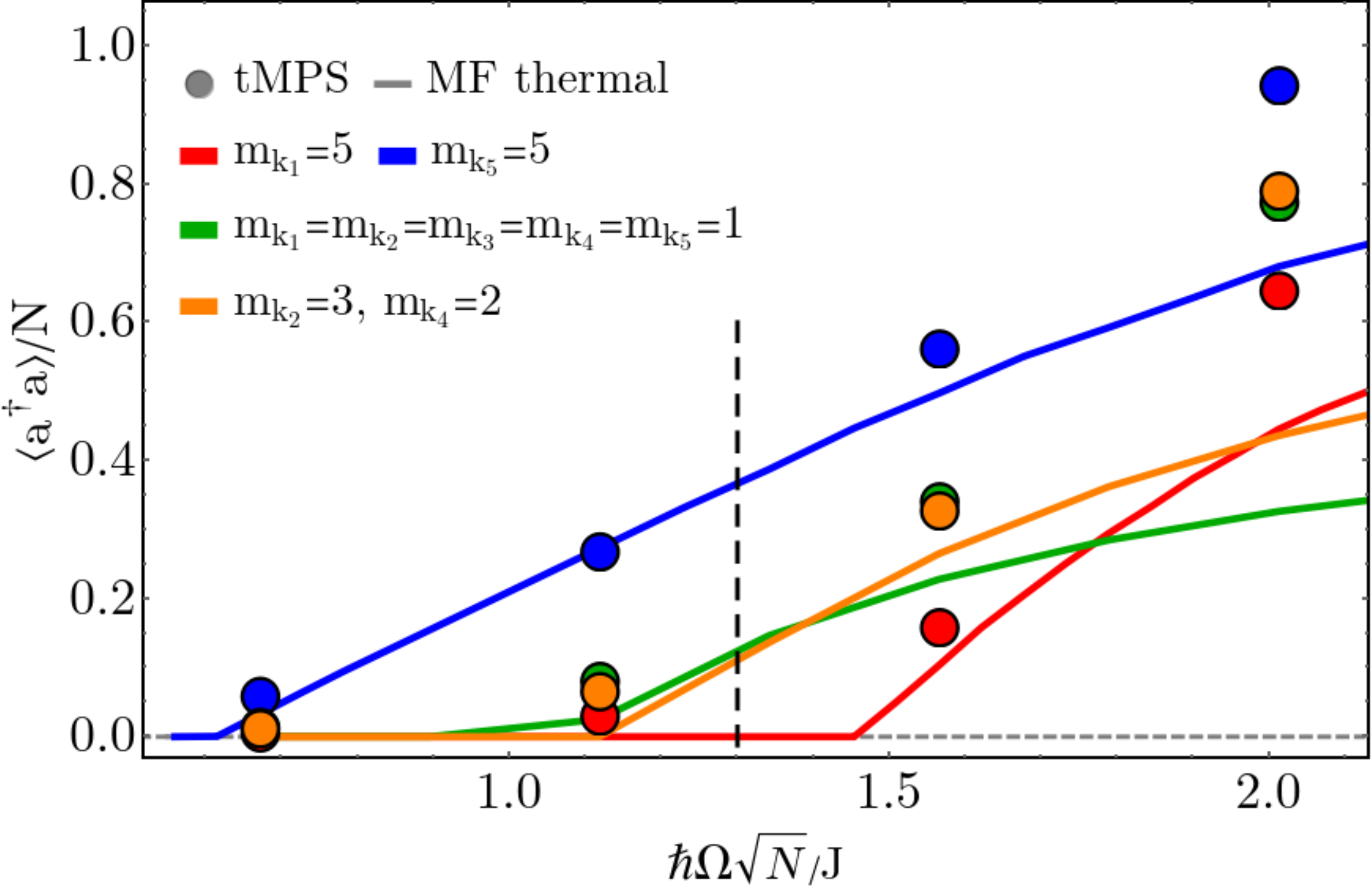}
\caption{The scaled photon number, $\langle a^\dagger a\rangle/N$, as a function of $\hbar\Omega\sqrt{N}/J$, for different symmetry sectors. We compare the values obtained with time-dependent matrix product state (tMPS) at $t J/\hbar=200$ with the generalized Gibbs ensemble within the mean field approach including thermal fluctuations.
The parameters used are $L=10$, $N=5$, $\hbar\delta/J=2$, $U/J=0$, and $\hbar\Gamma/J=1$. At the dashed vertical line, the nature of the obtained steady states reaches from states with an empty cavity (red symbol/line) to the self-organized states (remaining ones). 
 }
\label{fig:transition}
\end{figure}

For $U=0$, only transitions between the occupation of the momenta $k_j$ and $k_j+\pi~(\mathrm{mod}~2\pi)$ of the atoms are possible. 
In the single particle case, $L/2$ independent symmetry sectors exist, each spanned by the momentum states $\ket{k_j}$ and $\ket{k_j+\pi~(\mathrm{mod}~2\pi)}$, $j=1,\dotsc,L/2$. These are corresponding to the strong symmetry, having as generator the atomic number operators in each symmetry sector,
$\mathcal{O}_{k_j}= b^\dagger_{k_j} b_{{k_j}}+b^\dagger_{k_j+\pi~(\mathrm{mod}~2\pi)} b_{k_j+\pi~(\mathrm{mod}~2\pi)}$,
and their average values are conserved quantities, $m_{k_j}=\langle\mathcal{O}_{k_j}\rangle$. Due to the strong symmetry, already for a single particle multiple steady states exist.

For $N$ atoms, the symmetry sectors can be constructed from the different combinations in which one can arrange the atoms in the single particle sectors.
Thus, each symmetry sector will be labeled by $K\equiv\left(m_{k_1},...,m_{k_i},...,m_{k_{L/2}}\right)$, with $\sum_{i=1}^{L/2} m_{k_i}=N$. However, even though the atoms can be arranged independently in the single particle sectors, they are coupled via the photon field.

\textit{Nature of the steady states}. We show in Fig.~\ref{fig:transition} the steady state diagram of coupled atom cavity system. We observe in each considered symmetry sector of the strong symmetry a transition from the empty cavity state to the self-organized state with finite cavity occupation. Importantly, the transitions in the different symmetry sectors take place at distinct critical values of the coupling strength $\Omega_c$. Physically, this can be understood in the following way: the self-organization transition arises due to a competition between the ordering of the atoms in a density wave induced by interaction with the photon mode and the kinetic energy of the atoms which depends on the momentum of the atoms. As in each symmetry sector the atoms have different momenta, this gives rise to different critical values for the transition. Thus, at a fixed coupling strength (cf.~vertical line in Fig.~\ref{fig:transition}) multiple steady states arise depending on the projection of the initial state to the symmetry sectors. For the considered strong symmetry, these steady states can even have very different nature. It can occur that one sector is still in the disordered phase with an empty cavity, whereas another sector already is deep in the self-organized phase. This is to be contrasted to the meta-stable states arising for weak symmetries, which are typically connected to a unique steady state.

\begin{figure}[!hbtp]
\centering
\includegraphics[width=.49\textwidth]{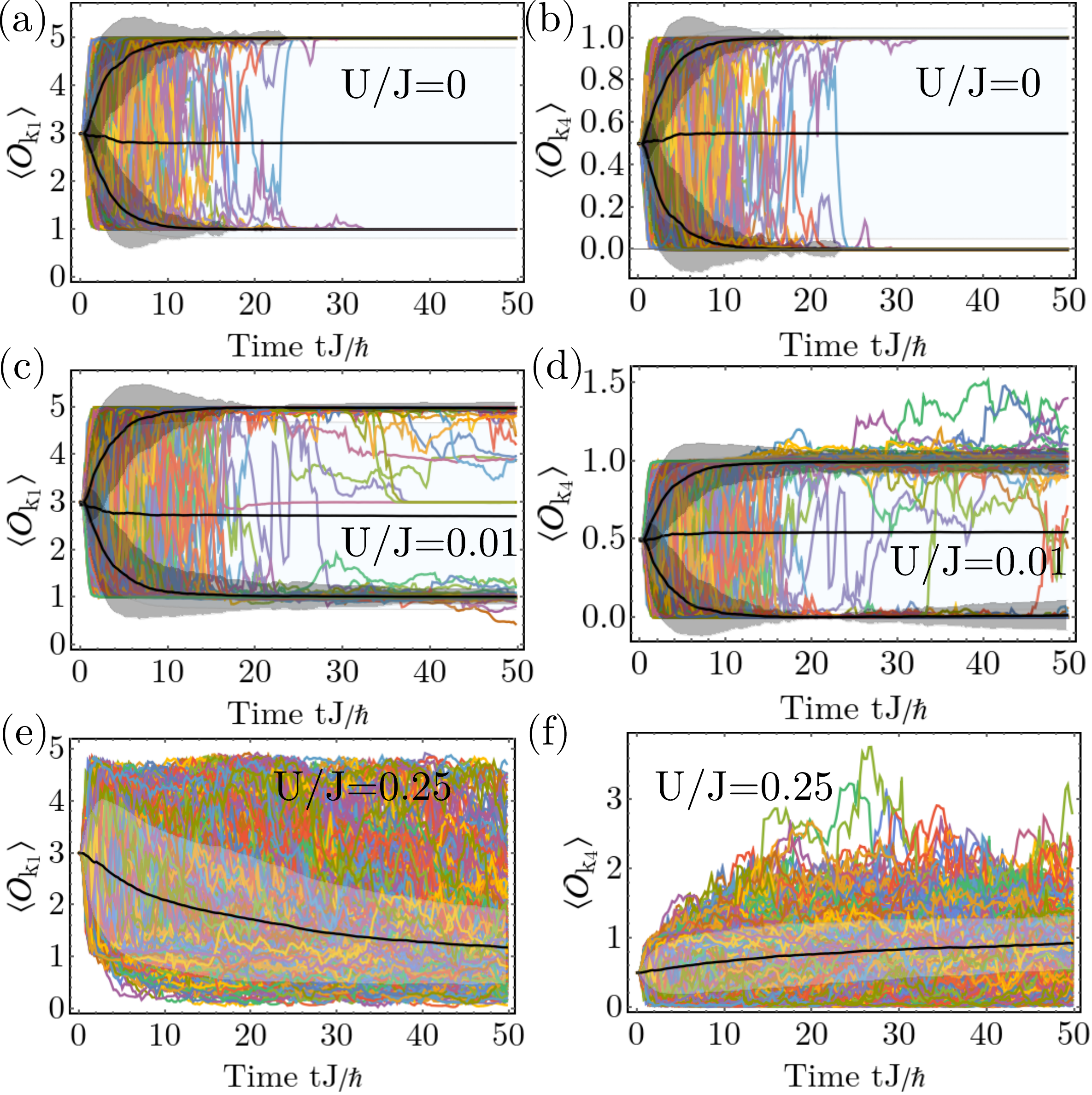}
\caption{ Time evolution of $\mathcal{O}_k$ for the single quantum trajectories sampled in the Monte Carlo average for different interaction strengths $U$, with $k=k_1$ (left column) and $k=k_4$ (right column). The initial state consists in an equal superposition between states for the sectors $\left(m_{k_1}=5\right)$ and $\left(m_{k_1}=1,m_{k_2}=1,m_{k_3}=1,m_{k_4}=1,m_{k_5}=1\right)$. In each panel there are 1000 trajectories plotted, the black represent the Monte Carlo average, either for the full set of trajectories, or averaged separately depending on the final value, we shade the interval of one standard deviation away from the average, with light blue for the full average and light gray for the separate averages. The parameters used are $L=10$, $N=5$, $\hbar\delta/J=2$, $\hbar\Omega\sqrt{N}/J=4.47$, and $\hbar\Gamma/J=15$.
 }
\label{fig:freezing}
\end{figure}

The results presented in Fig.~\ref{fig:transition} are obtained using two different methods. The first approach is a mean field decoupling of the atomic and the photonic sector considering the fluctuations around the mean-field solution as a perturbation together with the assumption that %within each symmetry sector
the atoms thermalize \cite{sup, BezvershenkoRosch2020}. 
In the presence of the strong symmetry we generalize this many-body adiabatic elimination approach by considering the conservation laws and using the thermalization of the atoms {\it within each} ($k$, $k+\pi~(\mathrm{mod}~2\pi)$)-sector with sector-dependent temperatures. This corresponds to a generalization of the 'generalized' Gibbs ensemble to dissipative systems \cite{sup}. We verify its applicability with an independent method. We expect that this generalization of the methods to be widely used in future to consider dissipative systems in the presence of symmetries. Our second approach is a matrix product state (MPS) method developed \cite{HalatiKollath2020b} for the numerically exact simulation of the time-evolution of the dissipative master equation, Eqs.~(\ref{eq:Lindblad})-(\ref{eq:Hamiltonian}) (the numerical parameters are given in \cite{sup}). For the time-evolution we have chosen the empty cavity and the ground state in the atomic sector as the initial state.

%\section{Dissipative freezing}

\textit{Dissipative freezing}. Recently, the effect of the existence of a strong symmetry in Liouvillian on the time evolution of the quantum trajectories was analyzed and the phenomenon of 'dissipative freezing' was shown for systems in which the Lindblad operator is proportional to the Hamiltonian ($H\propto L$) \cite{MunozPorras2019}. Dissipative freezing is the phenomenon that single realizations of trajectories, obtained by the stochastic unravelling of the master equation, can break the strong symmetry. A trajectory which is purely in one symmetry sector, will remain in that sector for the rest of the time evolution and thus, obey the symmetry of the system. However, starting with an initial state which is a superposition with contributions from multiple symmetry sectors, each individual trajectory will randomly select one of the sectors and remain there for the rest of the evolution. Further, no intermittency occurs between these trajectories in different sectors. Thus, even though the Monte Carlo average expectation value of the generator of the symmetry is a conserved quantity, this is no longer true at the level of single trajectories. The single trajectories can break 'spontaneously' the strong symmetry of the model. This is an intriguing effect which might have relevance to the single realizations of experiments. Further, the interpretation of quantities measured in the quantum trajectory method which stabilize at different values, should not imply different steady states, but reflect the initial superposition.

Here we numerically show [Fig.~\ref{fig:freezing}(a)-(b)] that even in a system that goes beyond the special case ($H\propto L$) of Ref.~\cite{MunozPorras2019}, dissipative freezing can occur. We show the evolution of the initial state which is an equal superposition of a state from the sector $(m_{k_1}=5)$ and the sector $(m_{k_1}=1, m_{k_2}=1, m_{k_3}=1, m_{k_4}=1, m_{k_5}=1)$. We can observe that at long times, $tJ\gtrsim 40\hbar$, all trajectories evolved to one of the two symmetry sectors, as $\langle\mathcal{O}_k\rangle(t)$ equals the expected occupation in those sectors. The Monte Carlo average of the trajectories stays constant throughout the following time-evolution, up to a numerical error.
Figs.~\ref{fig:freezing}(c)-(f) correspond to a finite on-site interaction and will be discussed later.

%\section{Breaking of the strong symmetry}

\textit{Breaking of the strong symmetry}. For any finite interaction, $U>0$, the operators $\mathcal{O}_k$, no longer commute with the Hamiltonian, thus the strong symmetry of the Liouvillian is broken. We analyze how the system passes over from having multiple steady states to a unique steady state as the on-site interaction is slowly turned on. We focus on the limit of large dissipation, for which the generalized many-body adiabatic elimination predicts the steady state transition between the multiple steady states at $U=0$, $\rho_{K,\text{st}}$ \cite{sup} to a single steady state which is the totally mixed state, $\rho_{\text{mix}}$ \cite{sup,HalatiKollath2020b}.

\begin{figure}[hbtp]
\centering
\includegraphics[width=.49\textwidth]{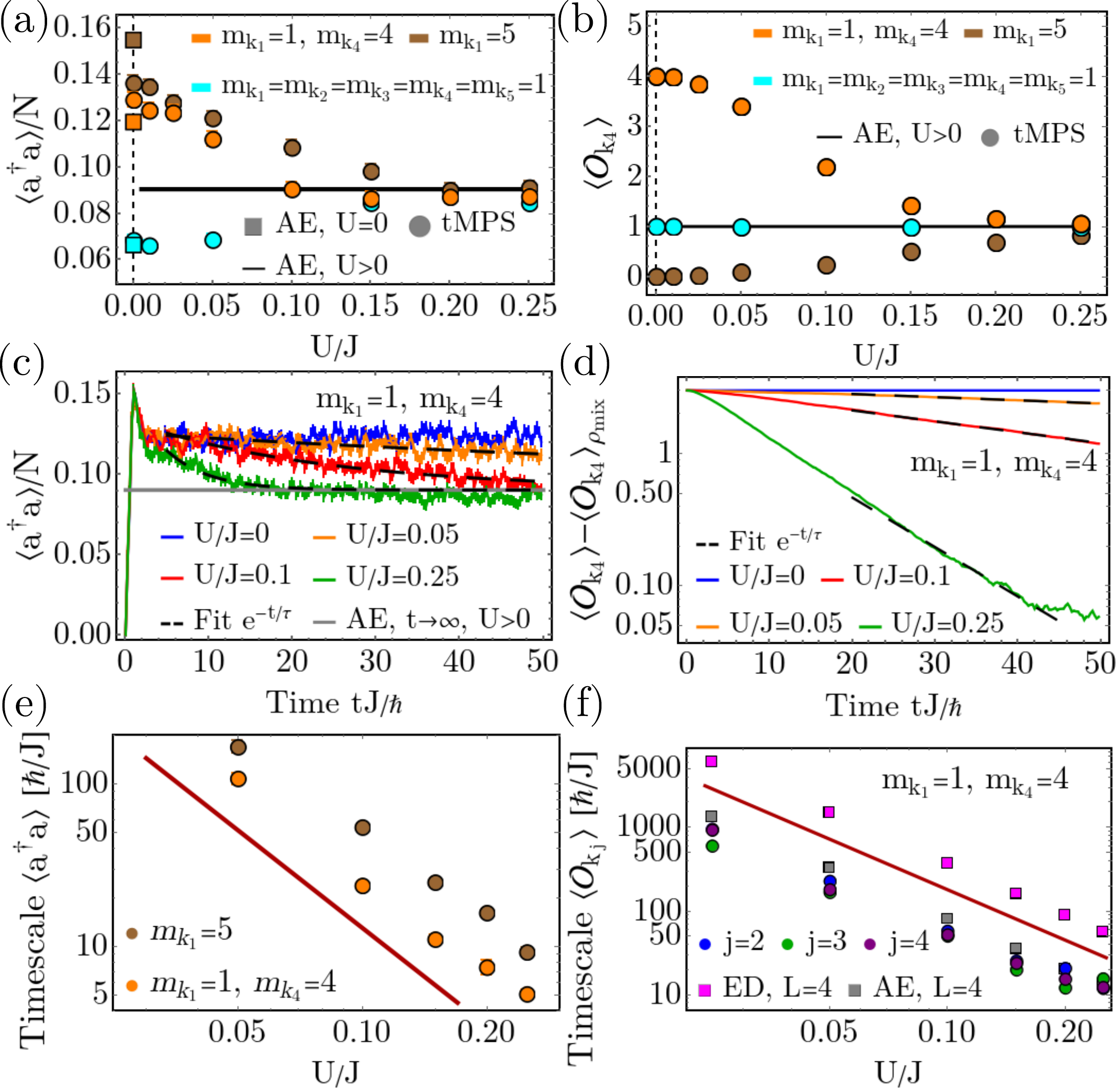}
\caption{The dependence on the interaction strength $U$ of (a) the scaled photon number, $\langle a^\dagger a\rangle/N$, and (b) the expectation value of $\mathcal{O}_{k_4}$ using tMPS at time $tJ=49.75\hbar$ and many-body adiabatic elimination (AE). 
The time evolution of (c) the scaled photon number, $\langle a^\dagger a\rangle/N$, and (d) the expectation value of $\mathcal{O}_{k_4}$ for different values of $U$. For finite $U$ we fit the time evolution with an exponential decay (black dashed lines) the difference between the tMPS data and the expected steady state value, obtained from many-body adiabatic elimination, using the kinetic energy as a perturbation.
The timescales obtained from the exponential fits are plotted a function of $U$ in log-log scale for (e) $\langle a^\dagger a\rangle/N$ and (f) $\mathcal{O}_{k_j}$. In (f) we compare the timescales of $\mathcal{O}_{k_j}$ with the longest timescale of a small system of $N=2$ particles in $L=4$ system computed with exact diagonalization (ED) and many-body adiabatic elimination (AE).
We fit the timescale dependence on the interaction with an algebraic decay $\propto U^{-\alpha}$ and obtain the following exponents: (e) $\left(m_{k_1}=5\right)$, $\alpha=1.86 \pm 0.07$; $\left(m_{k_1}=1, m_{k_4}=4\right)$, $\alpha=1.73 \pm 0.08$; (f) $j=2$, $\alpha=1.92 \pm 0.06$; $j=3$, $\alpha=1.77 \pm 0.09$; $j=4$, $\alpha=1.78 \pm 0.03$.
The red lines are a guide to the eye of an algebraic decay $\propto U^{-2}$.
The parameters are chosen to be $L=10$, $N=5$, $\hbar\Omega\sqrt{N}/J=4.47$, $\hbar\delta/J=2$, and $\Gamma/J=15$.  
 }
\label{fig:ae}
\end{figure}

In Fig.~\ref{fig:ae}(a)-(b) the behavior of the expectation value of the photon number and the conserved quantities of the symmetry at fixed time $tJ=49.75\hbar$ is plotted as a function of the interaction strength. We consider initial states in different symmetry sectors. For these, we observe that at $U=0$ multiple steady states are obtained signaled by distinct expectation values. 
However, as the interaction strength is increased the values of the photon number and $\langle\mathcal{O}_k\rangle$ for the different initial states start to be more and more similar until they agree with each other and with the values expected for $\rho_{\text{mix}}$, for large values of interaction $U$. The deviations from the predicted unique steady state $\rho_{\text{mix}}$ for small interaction strength are due to the fact that the system has not yet reached its steady state at the shown time. This can be observed in time-evolution plots given in Fig.~\ref{fig:ae}(c)-(d) for the photon number and $\langle\mathcal{O}_k\rangle-\langle\mathcal{O}_k\rangle_{\rho_{\text{mix}}}$ for different interaction values. The expected steady state value for $\langle a^\dagger a\rangle/N$ is represented with a gray line in Fig.~\ref{fig:ae}(c)and $\langle\mathcal{O}_k\rangle_{\rho_{\text{mix}}}=1$. 
In order to quantify this we fitted the time dependence of $\langle a^\dagger a\rangle-\langle a^\dagger a\rangle_{\rho_{\text{mix}}}$ and $\langle\mathcal{O}_k\rangle-\langle\mathcal{O}_k\rangle_{\rho_{\text{mix}}}$ with an exponential function, $\propto e^{-t/\tau}$, and extracted the timescales for reaching the steady state. The fits describe very nicely the numerical data, which gives strong support that at infinite time the steady state is given by $\rho_{\text{mix}}$. Additionally, 
the dependence of the timescale on $U$ is represented in Fig.~\ref{fig:ae}(e)-(f) in log-log plots. The timescales exhibits an algebraic dependence on $1/U^2$. 
We compare this behavior with results from the exact diagonalization of either the full Liouvillian, Eq.~(\ref{eq:Lindblad}), (ED) or the many-body adiabatic elimination equations of motion, with the kinetic energy as a perturbation \cite{HalatiKollath2020b}, (AE), for a small system of $L=4$. 
In these case we compute the timescale as the inverse of the real part of the first excited eigenvalue and we obtain an algebraic dependence $\propto U^{-\alpha}$ with $\alpha \approx 2$ confirming the tMPS results. We observe that the timescales for $\langle\mathcal{O}_k\rangle$ are consistently larger than the timescales for the photon number, which signals that in our simulations the photon state is reaching the steady state before the atomic one.  We attribute this to the spatial extent of the atomic system. 

Thus, we see that the time-evolution at short times remembers well the strong symmetry and the mixing of the different symmetry sectors only occurs on timescales $\propto 1/U^2$ associated with the scattering of the atoms. 

A question which arises is how the breaking of the symmetry affects the phenomenon of dissipative freezing discussed before. We study in the following how this phenomenon is affected by the presence of a small interaction, the symmetry breaking term.

We observe in Figs.~\ref{fig:freezing}(c)-(d) that at $U/J=0.01$ the time evolution found for the single trajectories resembles at early time the one at $U/J=0$. This means that the single trajectories break the approximate strong symmetry and approach the two different symmetry sectors.
At intermediate time, $20\lesssim tJ/\hbar$, many of the quantum trajectories spend a long time near the two values expected from the $U=0$ symmetry sectors. Only few of the trajectories directly show deviations from these values or intermittency between the values such that the phenomenon of 'dissipative freezing' also occurs here to an approximate extent. One has to be careful not to misinterpret this absence of intermittency as the existence of multiple steady state. 
Only starting from $U/J\gtrsim 0.25$ the effect of the strong symmetry  washes out in the considered time interval.

%\section{Conclusion}
To summarize, we analyzed the effects of a strong symmetry and of slightly breaking of this symmetry on the dynamics of a many-body open system consisting of bosonic atoms coupled to an optical cavity. The strong symmetry stems from the comensurability of the one-dimensional atomic chain with the cavity field.
Such a strong symmetry implies the existence of multiple steady states and we showed that the dissipative phase transition to the self-organized state can occur at different thresholds in different symmetry sectors described by generalized Gibbs ensembles in the atomic part. We analyzed how the nature of the steady state changes drastically when a small term that breaks the strong symmetry is introduced. The  timescales towards the new unique steady state where found to be proportional to $1/U^2$ associated with the scattering between the atoms.
We have shown that even for a many-body system with a strong symmetry the phenomenon of dissipative freezing can occur when one considers the behavior of individual quantum trajectories. It appears such that, at intermediate time, one can still identify the effect of dissipative freezing even if the strong symmetry has been slightly broken.
An open question remains, whether a spontaneous symmetry breaking can also be observed in single trajectories of an experimental measurement. This would question the interpretation of the absence of intermittency in experimental measurements.

\textit{Acknowledgments:}
 We thank J.-S. Bernier, A. V. Bezvershenko, M. Fleischhauer, A. Rosch and S. Wolff for stimulating discussions. We acknowledge funding from the Deutsche Forschungsgemeinschaft (DFG, German Research Foundation) in particular under project number 277625399 - TRR 185 (B3) and project number 277146847 - CRC 1238 (C05) and under Germany’s Excellence Strategy – Cluster of Excellence Matter and Light for Quantum Computing (ML4Q) EXC 2004/1 – 390534769 and the European Research Council (ERC) under the Horizon 2020 research and innovation programme, grant agreement No. 648166 (Phonton).

\clearpage

\section*{Supplemental Material}

\setcounter{section}{0}
\renewcommand{\thesection}{\Alph{section}}
\setcounter{equation}{0}
\renewcommand{\theequation}{A.\arabic{equation}}

\section{\label{appA}Open boundary conditions}

In the main text the Liouvillian describing the system, Eqs.(1)-(2), was given in momentum space, in real space the Hamiltonian of the system reads

\begin{align} 
\label{eq:Hamapp}
&H_0=H_c+H_{\text{kin}}+H_{\text{ac}}, \\
&H_c= \hbar\delta a^\dagger a\nonumber,\\
&H_{\text{kin}}=-J \sum_{j=1}^{L-1} (b_{j}^\dagger b_{j+1} + b_{j+1}^\dagger b_{j}),\nonumber\\
&H_{\text{ac}}=  -\hbar\Omega ( a + a^\dagger) \sum_{j=1}^L (-1)^j n_j, \nonumber\\
&H=H_0+H_{\text{int}}, \nonumber\\
&H_{\text{int}}=\frac{U}{2} \sum_{j=1}^L n_{j}(n_{j}-1).\nonumber
\end{align}

In our numerical simulations for finite size systems we used open boundary conditions. In this case one uses the Fourier sine transform, defined by

\begin{align} 
\label{eq:fourier}
b_k &=\sqrt{\frac{2}{L+1}}\sum_{j=1}^{L} b_j \sin(kj),\\
b_k^\dagger &=\sqrt{\frac{2}{L+1}}\sum_{j=1}^{L} b_j^\dagger \sin(kj), \nonumber
\end{align}
and the unitless momenta are given by $k=\frac{\pi m}{L+1}$ and $m=1,\dotsc,L$. In this case still $L/2$ independent symmetry sectors exist for a single particle, but each is spanned by the momentum states $\ket{k_j}$ and $\ket{\pi-k_j}$, $j=1,\dotsc,L/2$. As the momenta $k_j$ are in the interval $[0,\pi]$, the values $\pi-k_j$ will always be inside the first Brillouin zone.
The symmetry generator are now given by
\begin{align} 
\label{eq:genmomobc}
\mathcal{O}_{k_j}&= b^\dagger_{k_j} b_{{k_j}}+b^\dagger_{\pi-{k_j}} b_{\pi-{k_j}}. 
\end{align}
We note that in the calculations presented in Sec.~\ref{appB} we employ open boundary conditions. 

\setcounter{equation}{0}
\renewcommand{\theequation}{B.\arabic{equation}}
\setcounter{figure}{0}
\renewcommand{\thefigure}{B\arabic{figure}}

\section{\label{appB}Generalized many body adiabatic elimination formalism}

We employ a variant of the many-body adiabatic elimination method \cite{Garcia-RipollCirac2009, ReiterSorensen2012, PolettiKollath2013, SciollaKollath2015}, which provides analytical insight into the long-time behavior of our system. This approach is a perturbative approach which considers that the effect of one of the terms in the Hamiltonian, $H_{\nu}$, or of the fluctuations around the mean-field solution in the dynamics of the system is weak.
We used this procedure in Ref.~\cite{HalatiKollath2020, HalatiKollath2020b, BezvershenkoRosch2020} to determine the phase diagram of steady states in the presence of interactions.

We decompose the Liouvillian as $\mathcal{L}=\mathcal{L}_0-\frac{i}{\hbar}[H_{\nu},\cdot]$ into  an unperturbed Linbladian $\mathcal{L}_0$ and a perturbative contribution caused by $H_\nu$. 
This approach captures the effective dynamics of the density matrix in the decoherence free subspace of $\mathcal{L}_0$, i.e.~the space formed by all density matrices $\rho_0$ which are eigenstates of the superoperator $\mathcal{L}_0$ with a vanishing real part of the eigenvalues. The other subspaces corresponding to the eigenvalues with a non-zero real part are considered via virtual transitions within the perturbation theory. The resulting effective dynamics in the decoherence free subspace is given by \cite{Garcia-RipollCirac2009,PolettiKollath2013, SciollaKollath2015, HalatiKollath2020b}
\begin{align}
\label{eq:decfree0}
\frac{d}{dt}\rho^{0} =\mathcal{L}_0 \rho^0+\frac{1}{\hbar^2} P_0 \left[ H_{\nu},\mathcal{L}_0^{-1} P_1 \left[H_{\nu},\rho^0\right]\right],
\end{align}
where $\rho^0$ lies in the decoherence free subspace of $\mathcal{L}_0$ and $P_0$ and $P_1$ are the projectors onto the decoherence free subspace and the first excited subspace, respectively. In the following we detail how to apply this approach. In Sec.~\ref{sec:meanfield} the perturbation is the fluctuations around the mean-field theory and in Sec.~\ref{sec:kinetic} the perturbation is the kinetic term of the atoms. 

We note that due to the applied perturbative expansion the condition of positive definiteness might not be fulfilled for the obtained density matrix \cite {LiKoch2014}. 

\renewcommand{\thesubsection}{\Roman{subsection}}

\subsection{Mean field decoupling with thermal fluctuations}
 \label{sec:meanfield}
Following the approach introduced in Ref.~\cite{BezvershenkoRosch2020} we perform a mean field decoupling of the term coupling the cavity and the atoms, $H_\text{ac}$, and consider the fluctuations in the coupling as the perturbation in the many body adiabatic elimination derivation of the effective equations of motion. In this situation we have

\begin{align}
\label{eq:lio0mf}
 \mathcal{L}_0 &=-\frac{i}{\hbar}[H_c+H_{\text{kin}}+H_{\text{ac}}^\text{MF},\cdot]+\mathcal{D}(\cdot), \\
H_{\nu} &\equiv \delta H_{\text{ac}} \nonumber
 \end{align}
where $\mathcal{D}(\rho)=\frac{\Gamma}{2}\left(2a\rho a^\dagger-a^\dagger a \rho-\rho a^\dagger a\right)$, and  
\begin{align}
\label{eq:mfdec}
H_{\text{ac}}^\text{MF} =& -\hbar \Omega (\alpha+\alpha^*) \sum_{k} b_{k}^\dagger b_{\pi-k} \\
&\quad-\hbar \Omega ( a + a^\dagger) \Delta, \nonumber \\
\delta H_{\text{ac}} =& -\hbar \Omega\left( a + a^\dagger-\alpha-\alpha^*\right)\left[\sum_{k} b_{k}^\dagger b_{\pi-k}- \Delta\right], \nonumber.
\end{align}
Here $\alpha$ is the mean field value of $\langle a\rangle$ which depends on the mean field value of the imbalance $\Delta$,
\begin{align}
\label{eq:mfsc}
 &\alpha(\Delta) =\frac{\Omega}{\delta-i\Gamma/2} \Delta, \\
&\Delta=\sum_{j=1}^{L/2}\langle\mathcal{O}_{k_j}\rangle=\sum_{j} (-1)^j \langle n_j \rangle. \nonumber
\end{align}

Here we see that the atomic and photonic contributions decouple beside the self-consistent determination of the parameters. Thus, a general state in the decoherence free subspace of $\mathcal{L}_0$ is given by 
\begin{align}
\label{eq:rhoL0}
   &\rho= |\alpha (\Delta) \rangle\langle \alpha (\Delta)|\cdot\rho^b,~\text{with}~\rho^b=\sum_{n,m} c_{n,m}\ket{n(\alpha)}\bra{m(\alpha)},
 \end{align}
where $\ket{n(\alpha)}$ are eigenstates of the atomic mean field Hamiltonian $H_a=H_{\text{kin}}+H_{\text{ac}}^\text{MF}$ and the cavity part is in a coherent state. If we plug in $\rho$ given by Eq.~(\ref{eq:rhoL0}) in Eq.~(\ref{eq:decfree0}) we obtain the equation of motion for the entries $c_{n,m}$ of the density matrix \cite{HalatiPhD}. A substantial simplification can occur if we consider a generalization of the thermal Ansatz introduced in Ref.~\cite{BezvershenkoRosch2020}. As we only have cavity mediated interactions we cannot assume that the atomic sector thermalizes as a whole, but we can make the assumption that the particles in each single particle symmetry sector can thermalize and be described by an effective inverse temperature $\beta_j$ associated with this sector, i.e.
\begin{align}
\label{eq:rhobeta}
   &\qquad\rho^b \sim \prod_{j=1}^{L/2} \exp \left[-\beta_j H_a(k_j,\pi-k_j)\right], ~~\text{where} \\
   &H_a(k_j,\pi-k_j)=-\hbar\Omega (\alpha+\alpha^*) \left(b_{k_j}^\dagger b_{\pi-k_j}+b_{\pi-k_j}^\dagger b_{k_j}\right) \nonumber\\
   &-2J \cos(k_j) \left(b_{k_j}^\dagger b_{k_j}-b_{\pi-k_j}^\dagger b_{\pi-k_j}\right). \nonumber
\end{align}
For example, in Fig.~1, the state from the symmetry sector $\left(m_{k_1}=5\right)$ is described by a single temperature as all particles are in the single particle sector with momentum $k_1$, but the state from the symmetry sector $\left(m_{k_2}=3,m_{k_4}=2 \right)$ is described by two temperatures as we have 3 particles in the single particle sector with momentum $k_2$ and 2 particles in the single particle sector with momentum $k_4$.
This procedure is analogously to the consideration of different conservation laws in closed systems, described by generalized Gibbs ensembles. Let us note that a generalized Gibbs ensemble was also introduced in Ref.~\cite{LenarcicRosch2018} for weakly driven systems in the presence of approximate conservation laws. 

As the density matrix is now determined by a smaller number of parameters, the inverse temperatures $\beta_j$, it is enough to consider the equations of motion for a reduced number of observables. Thus, we describe the steady state of the system with the temperatures for which the Ansatz given by Eqs.~(\ref{eq:rhoL0})-(\ref{eq:rhobeta}) satisfies the equations $\left\langle \pdv{t} H_a(k_j,\pi-k_j)\right\rangle=0$ for all momenta $k_j$ and the mean-field self-consistency condition Eq.~(\ref{eq:mfsc}).

\subsection{Perturbation in kinetic energy}
 \label{sec:kinetic}

In the following, we consider the perturbation to be given by the kinetic energy, $H_{\nu}\equiv H_{\text{kin}}$, valid in the regime $\hbar\Gamma\gg \hbar\Omega, \hbar\delta \gg J$, thus $ \mathcal{L}_0$ is given by
\begin{align}
\label{eq:lio0}
 \mathcal{L}_0=-\frac{i}{\hbar}[H_c+H_{\text{int}}+H_{\text{ac}},\cdot]+\mathcal{D}(\cdot),
 \end{align}
with $\mathcal{D}(\rho)=\frac{\Gamma}{2}\left(2a\rho a^\dagger-a^\dagger a \rho-\rho a^\dagger a\right)$ the dissipator.

We start by constructing the dissipation free subspace and the excited subspaces of $\mathcal{L}_0$ which will enter our calculations, for an arbitrary interaction strength $U\geq 0$ \cite{HalatiKollath2020b}. Afterwards, we obtain the steady states for the both cases with finite interaction and without interactions.

The following states of the form
\begin{align}
\label{eq:ansatz_u}
& \rho=\ket{\alpha(\Delta);\Delta,u}\bra{\alpha(\Delta');\Delta',u'}
\end{align}
are right eigenstates of the superoperator $\mathcal{L}_0$. We note that we do not assure that these states are physical density matrices.
The atomic part is characterized by the even-odd imbalance $\Delta=\sum_{j} (-1)^j \langle n_j \rangle$ and by its total interaction energy, $u= \frac{U}{2}\sum_{j}\langle n_j(n_j-1) \rangle$. Photons are in a coherent state depending on the atomic imbalance
\begin{align}
\label{eq:coherent}
 \alpha(\Delta) &=\frac{\Omega}{\delta-i\Gamma/2} \Delta.
\end{align}

The eigenvalues corresponding to the eigenvectors from Eq.~(\ref{eq:ansatz_u}) are given by 
\begin{align}
\label{eq:eigenvalue_u}
\lambda =&  -\frac{1}{2}\frac{\Omega^2\Gamma}{\delta^2+\Gamma^2/4}(\Delta-\Delta')^2 \\
&+i\left[\frac{\Omega^2\delta}{\delta^2+\Gamma^2/4}(\Delta^2-\Delta'^2)-(u-u')\right]. \nonumber
\end{align}
We can observe that for $\Delta=\Delta'$ the real part of the eigenvalues is zero. Thus, the states in Eq.~(\ref{eq:ansatz_u}) with $\Delta=\Delta'$ lie in the decoherence free subspace of $\mathcal{L}_0$. 

We can include the contributions from the excited subspaces that are coupled to the dissipation free subspace via the perturbation $H_{\text{kin}}$. 
With a hopping event we can couple to the subspace spanned by the states in which $\Delta=\Delta'\pm 2$. 

In the case with finite interactions, one can explicitly write the equations of motion, Eq.~(\ref{eq:decfree0}) for the elements of the decoherence free subspace for the general case of $N$ particles in $L$ sites and can show that the mixed state given by \cite{HalatiKollath2020b} 
\begin{align}
\label{eq:ss_gen}
\rho_{\text{mix}} &=\frac{1}{\mathcal{N}}\sum_{\{n_j\}} \ket{\alpha(\Delta);n_1,\dotso,n_L}\bra{\alpha(\Delta);n_1,\dotso,n_L}
\end{align}
is a steady state of the system. Here $\mathcal{N}$ is the number of ways one can arrange $N$ identical particles in $L$ sites, $\mathcal{N}=\begin{pmatrix} L+N-1 \\ N \end{pmatrix}$. 

In the rest of this section we deal with the case of non-interacting atoms,
\begin{align}
\label{eq:lio00}
 \mathcal{L}_0=-\frac{i}{\hbar}[H_c+H_{\text{ac}},\cdot]+\mathcal{D}(\cdot).
\end{align}
As we are in the non-interacting case with the momentum labeling the different symmetry sectors, we first compute the steady state of one particle in a certain symmetry sector and, afterwards, generalize this result to the case of $N$ particles.

\subsubsection{The single particle case}

For a single particle in the symmetry sector $K=(m_k=1)$, a general state in the dissipation free subspace restricted to this symmetry sector has the form
\begin{align} 
\label{eq:momstates1pgen}
\rho_k (b)&=\sum_{i,j~\text{odd}}\left(\frac{1}{2}+b\right)\sin(ki)\sin(kj)\ket{-\alpha;i}\bra{-\alpha;j} \\
&+\sum_{i,j~\text{even}}\left(\frac{1}{2}-b\right)\sin(ki)\sin(kj)\ket{\alpha;i}\bra{\alpha;j}, \nonumber
\end{align}
with $b$ a real parameter, $i$ and $j$ the positions of the particle, and $\alpha=\frac{\Omega}{\delta-i \Gamma/2}$ the cavity field.

The equation of motion for a state $\rho_0=\ket{\pm\alpha;i}\bra{\pm\alpha;j}$, with $i$ and $j$ both even or both odd, from the dissipation free subspace, obtained with the many-body adiabatic elimination is given by \cite{HalatiPhD}
\begin{widetext}
\begin{align} 
\label{eq:momstates1peq}
\frac{d}{dt}\ket{\pm\alpha;i}\bra{\pm\alpha;j}=P_0 &\left[H_\text{kin},\mathcal{L}_0^{-1}P_1\left[H_\text{kin},\ket{\pm\alpha;i}\bra{\pm\alpha;j}\right]\right] \\
=\frac{J^2}{\lambda_0}e^{-4|\alpha|^2}&\Big(4\ket{\pm\alpha;i}\bra{\pm\alpha;j}-2\ket{\mp\alpha;i+1}\bra{\mp\alpha;j+1}-2\ket{\mp\alpha;i+1}\bra{\mp\alpha;j-1} \nonumber\\
&-2\ket{\mp\alpha;i-1}\bra{\mp\alpha;j+1}-2\ket{\mp\alpha;i-1}\bra{\mp\alpha;j-1}+\ket{\pm\alpha;i+2}\bra{\pm\alpha;j} \nonumber\\
&+\ket{\pm\alpha;i-2}\bra{\pm\alpha;j}+\ket{\pm\alpha;i}\bra{\pm\alpha;j+2}+\ket{\pm\alpha;i}\bra{\pm\alpha;j-2}\Big), \nonumber
\end{align}
\end{widetext}
with $\lambda_0=-\frac{2\Omega^2\Gamma}{\delta^2+\Gamma^2/4}$. From this we can write the equation of motion for the state $\rho_k (b)$, and for $b=0$ we obtain the steady state for the one particle case

\begin{align} 
\label{eq:momstates1pss}
\rho_{k,\text{st}}&=\sum_{i,j~\text{odd}}\sin(ki)\sin(kj)\ket{-\alpha;i}\bra{-\alpha;j} \\
&+\sum_{i,j~\text{even}}\sin(ki)\sin(kj)\ket{\alpha;i}\bra{\alpha;j}. \nonumber
\end{align}
This state has a fully mixed atomic sector in the momentum basis.

\subsubsection{The two particle case}

We consider two particles in the sector $K=(m_{k_1}=1,m_{k_2}=1)$. 
One can determine the basis in the dissipation free subspace in the considered symmetry sector and compute the equations of motion for these states \cite{HalatiPhD}.
From this one obtains the steady state solutions \cite{HalatiPhD}

\begin{widetext}
\begin{align} 
\label{eq:momss2p}
\rho_{k_1,k_2, \text{st}}=&\sum_{i_1,i_2,j_1,j_2~\text{odd}}\sin(k_1i_1)\sin(k_1j_1)\sin(k_2i_2)\sin(k_2j_2)\sqrt{n_{i_1}n_{j_1}}\ket{-\alpha;i_1,i_2}\bra{-\alpha;j_1,j_2} \\
&+\sum_{i_1,i_2,j_1,j_2~\text{even}}\sin(k_1i_1)\sin(k_1j_1)\sin(k_2i_2)\sin(k_2j_2)\sqrt{n_{i_1}n_{j_1}}\ket{\alpha;i_1,i_2}\bra{\alpha;j_1,j_2} \nonumber\\
&+\sum_{\substack{i_1,j_1~\text{odd} \\ i_2,j_2~\text{even}}}\sin(k_1i_1)\sin(k_1j_1)\sin(k_2i_2)\sin(k_2j_2)\sqrt{n_{i_1}n_{j_1}}\ket{0;i_1,i_2}\bra{0;j_1,j_2} \nonumber\\
&+\sum_{\substack{i_2,j_2~\text{odd} \\ i_1,j_1~\text{even}}}\sin(k_1i_1)\sin(k_1j_1)\sin(k_2i_2)\sin(k_2j_2)\sqrt{n_{i_1}n_{j_1}}\ket{0;i_1,i_2}\bra{0;j_1,j_2}, \nonumber
\end{align}
\end{widetext}
with $i_1$, $i_2$, $j_1$ and $j_2$ the positions of the two particles in the ket or bra, and $n_i$ the number of particles at site $i$, and $\alpha=\frac{2\Omega}{\delta-i \Gamma/2}$
If we trace out the photon states we recover a fully mixed atomic sector

\begin{align} 
\label{eq:momss2p}
\tr_\textrm{photons} \rho_{k_1,k_2, \text{st}}=&\frac{1}{4}(\ket{k_1,k_2}\bra{k_1,k_2}\\
&+\ket{k_1,\pi-k_2}\bra{k_1,\pi-k_2} \nonumber\\
&+\ket{\pi-k_1,k_2}\bra{\pi-k_1,k_2}\nonumber\\
&+\ket{\pi-k_1,\pi-k_2}\bra{\pi-k_1,\pi-k_2}). \nonumber
\end{align}
Thus, similar as in the case of the interacting system, in the limit of large dissipation and small kinetic energy, the infinite temperature state of the corresponding symmetry block is reached. 

\begin{figure*}[hbtp]
\centering
\includegraphics[width=.98\textwidth]{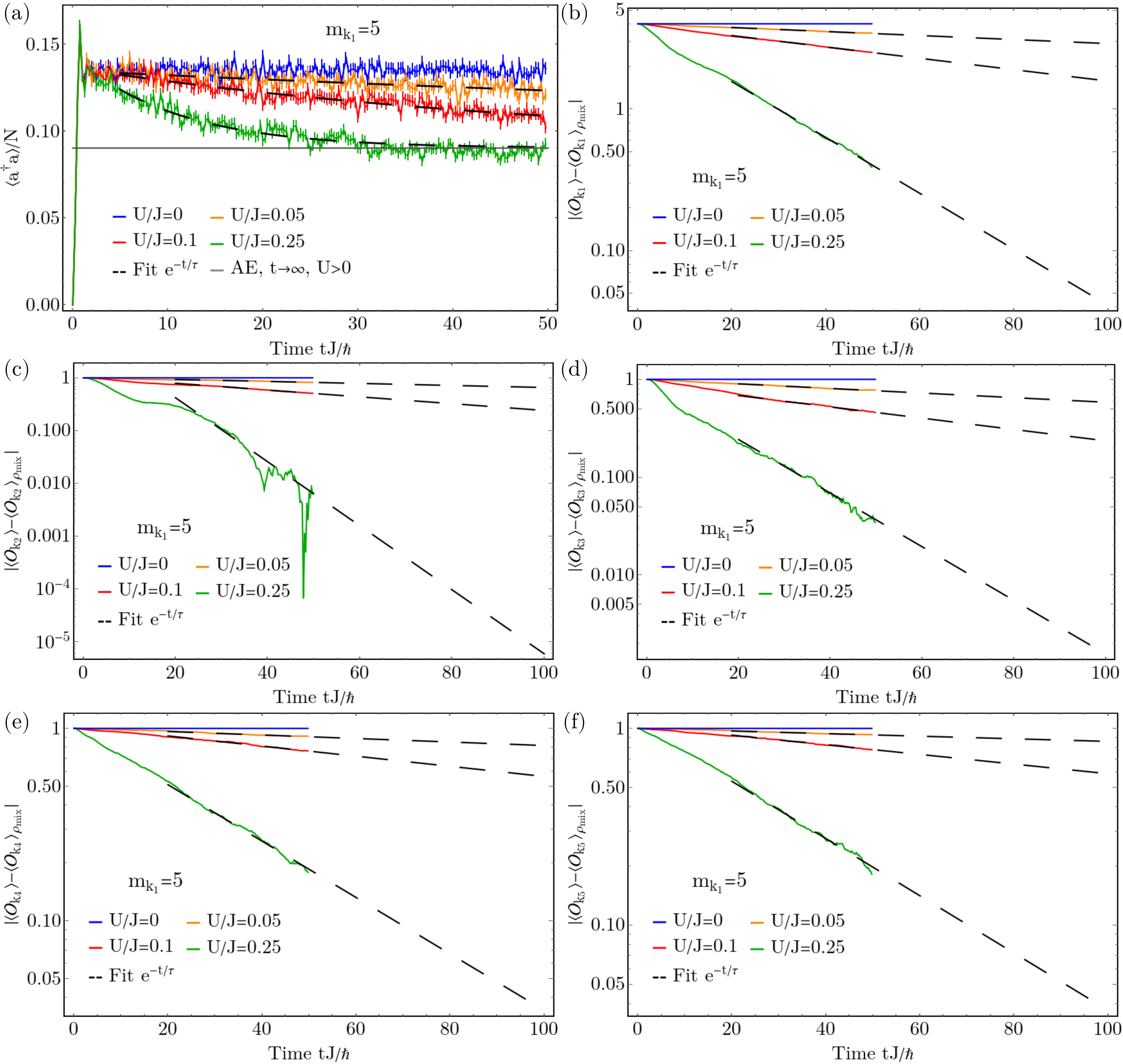}
\caption{The time evolution of (a) the scaled photon number, $\langle a^\dagger a\rangle/N$, and [(b)-(f)] the expectation value of $\mathcal{O}_{k_j}$ for different values of $U$. For finite $U$ we fit the time evolution with an exponential decay (black dashed lines) the difference between the tMPS data and the expected steady state value, obtained from many-body adiabatic elimination.
The parameters are chosen to be $L=10$, $N=5$, $\hbar\Omega\sqrt{N}/J=4.47$, $\hbar\delta/J=2$, $\Gamma/J=15$, and the symmetry sector $(m_{k_1}=5)$. 
}
\label{fig:breakingte1}
\end{figure*}

\begin{figure*}[hbtp]
\centering
\includegraphics[width=.98\textwidth]{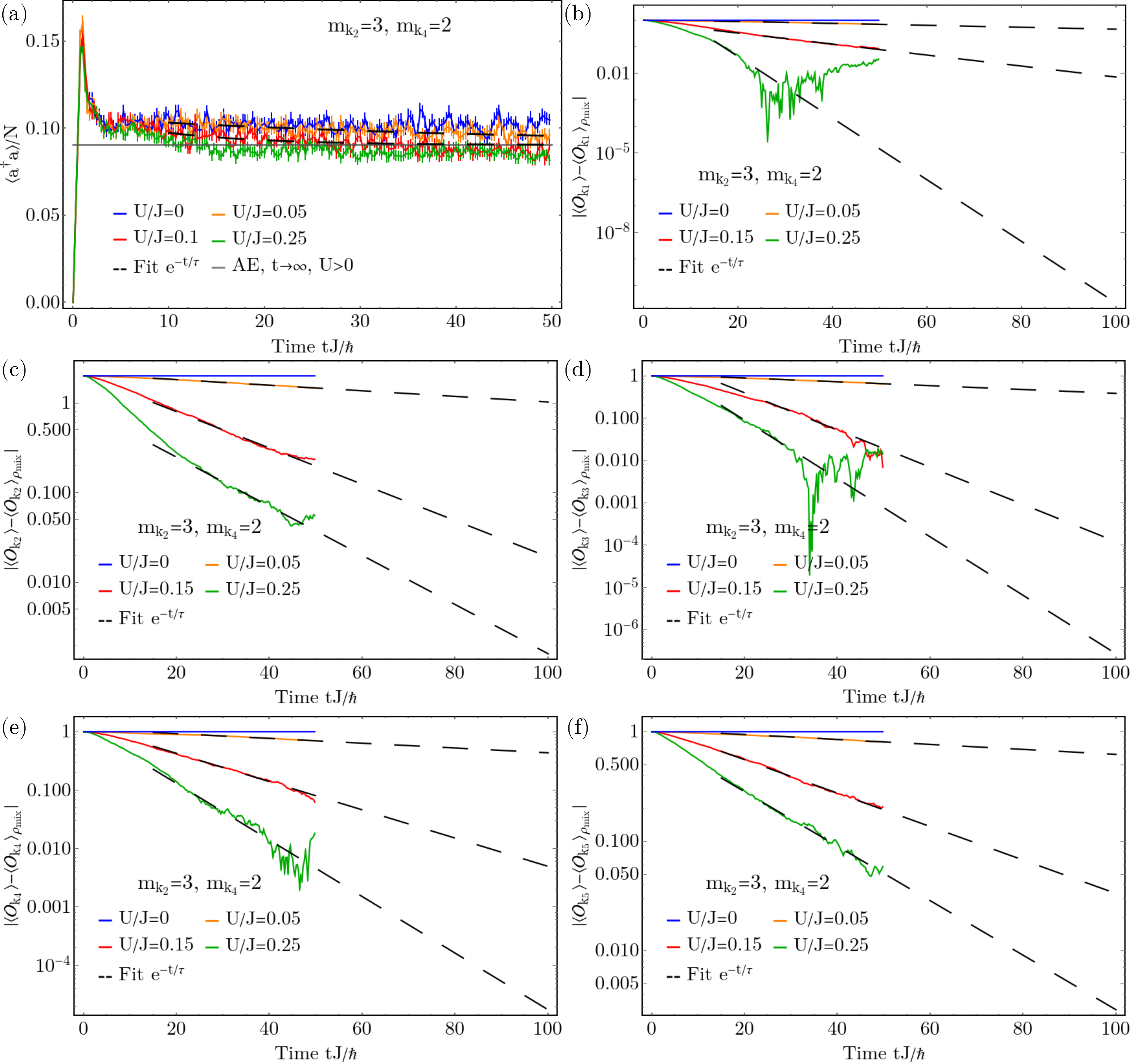}
\caption{The time evolution of (a) the scaled photon number, $\langle a^\dagger a\rangle/N$, and [(b)-(f)] the expectation value of $\mathcal{O}_{k_j}$ for different values of $U$. For finite $U$ we fit the time evolution with an exponential decay (black dashed lines) the difference between the tMPS data and the expected steady state value, obtained from many-body adiabatic elimination.
The parameters are chosen to be $L=10$, $N=5$, $\hbar\Omega\sqrt{N}/J=4.47$, $\hbar\delta/J=2$, $\Gamma/J=15$, and the symmetry sector $(m_{k_2}=3, m_{k_4}=2)$.   
}
\label{fig:breakingte2}
\end{figure*}

\subsubsection{$N$ particle case}
Generalizing our previous finding to the $N$ particle case, the steady state for $N$ particles will also be the fully mixed state in the different symmetry sectors. 

For the $N$ particle case the symmetry sectors can be constructed from different combinations in which one can arrange the particles in the single particle sectors. Thus, the totally mixed state for $N$ particles distributed in the $L/2$ single particle sectors, $K=\left(m_{k_1},...,m_{k_i},...,m_{k_{L/2}}\right)$, is given by

\begin{widetext}
\begin{align} 
\label{eq:momss}
&\rho_{\text{K,st}}=\frac{1}{\mathcal{N}_0}\sum_{i_1=0}^{m_{k_1}}\cdots\sum_{i_K=0}^{m_{k_L/2}}\ket{n_{k_1}=i_1,n_{\pi-k_1}=m_{k_1}-i_1;\cdots ;n_{k_{L/2}}=i_{L/2},n_{\pi-k_{L/2}}=m_{k_{L/2}}-i_{L/2}}\\
&\qquad\qquad\qquad\qquad\qquad\bra{n_{k_1}=i_1,n_{\pi-k_1}=m_{k_1}-i_1;\cdots ;n_{k_{L/2}}=i_{L/2},n_{\pi-k_{L/2}}=m_{k_{L/2}}-i_{L/2}},\nonumber \\
\text{with}\nonumber \\
&\ket{n_{k_1}=i_1,n_{\pi-k_1}=m_{k_1}-i_1;\cdots ;n_{k_{L/2}}=i_{L/2},n_{\pi-k_{L/2}}=m_{k_{L/2}}-i_{L/2}}\equiv  \nonumber\\
&\qquad\equiv\frac{1}{\mathcal{M}}\sum_{j_1^1,...,j^1_{m_{k_1}}=0}^{L}\cdots\sum_{j_1^{L/2},...,j^{L/2}_{m_{k_{L/2}}}=0}^{L} \left(\sin(k_1j_1^1)..\sin(k_1j_{m_{k_1}}^1)(-1)^{j_{i_1+1}+...+j_{m_{k_1}}+(m_{k_1}-i_1)}\right)\times\cdots \nonumber\\
&\qquad\qquad\qquad\qquad\qquad\qquad\qquad\quad\cdots\times  \left(\sin(k_{L/2}j_1^{L/2})..\sin(k_{L/2}j_{m_{k_{L/2}}}^{L/2})(-1)^{j_{i_{L/2}+1}+...+j_{m_{k_{L/2}}}+(m_{k_{L/2}}-i_{L/2})}\right)\times \nonumber\\
&\qquad\qquad\qquad\qquad\qquad\qquad\qquad\quad\times\sqrt{(n_1)!...(n_L)!}~\ket{\alpha(\Delta);j_1^1,...,j^1_{m_{k_1}},...,j_1^{L/2},...,j^{L/2}_{m_{k_{L/2}}}}, \nonumber
\end{align}
\end{widetext}
where $n_{k_i}$ is the number of particles with momentum $k_i$, $\mathcal{N}_0=\prod_{i=1}^{L/2}\begin{pmatrix} m_{k_i}+1 \\ m_{k_i} \end{pmatrix}$, $j_1^1,...,j^1_{m_{k_1}},...,j_1^{L/2},...,j^{L/2}_{m_{k_{L/2}}}$ are the positions of the $N$ particles and $n_i$ is the occupation number of each site in real space and the even-odd imbalance is given by $\Delta=\sum_{p=1}^{m_{k_1}}(-1)^{j^1_p}+...+\sum_{p=1}^{m_{k_{L/2}}}(-1)^{j^{L/2}_p}$, and the normalization constant is $\mathcal{M}=\left(\frac{L+1}{2}\right)^{N/2}\sqrt{(n_{k_1})!(n_{\pi-k_1})!...(n_{k_{L/2}})!(n_{\pi-k_{L/2}})!}$.

\subsubsection{Comparison with numerical exact tMPS results}

\begin{figure}[hbtp]
\centering
\includegraphics[width=.48\textwidth]{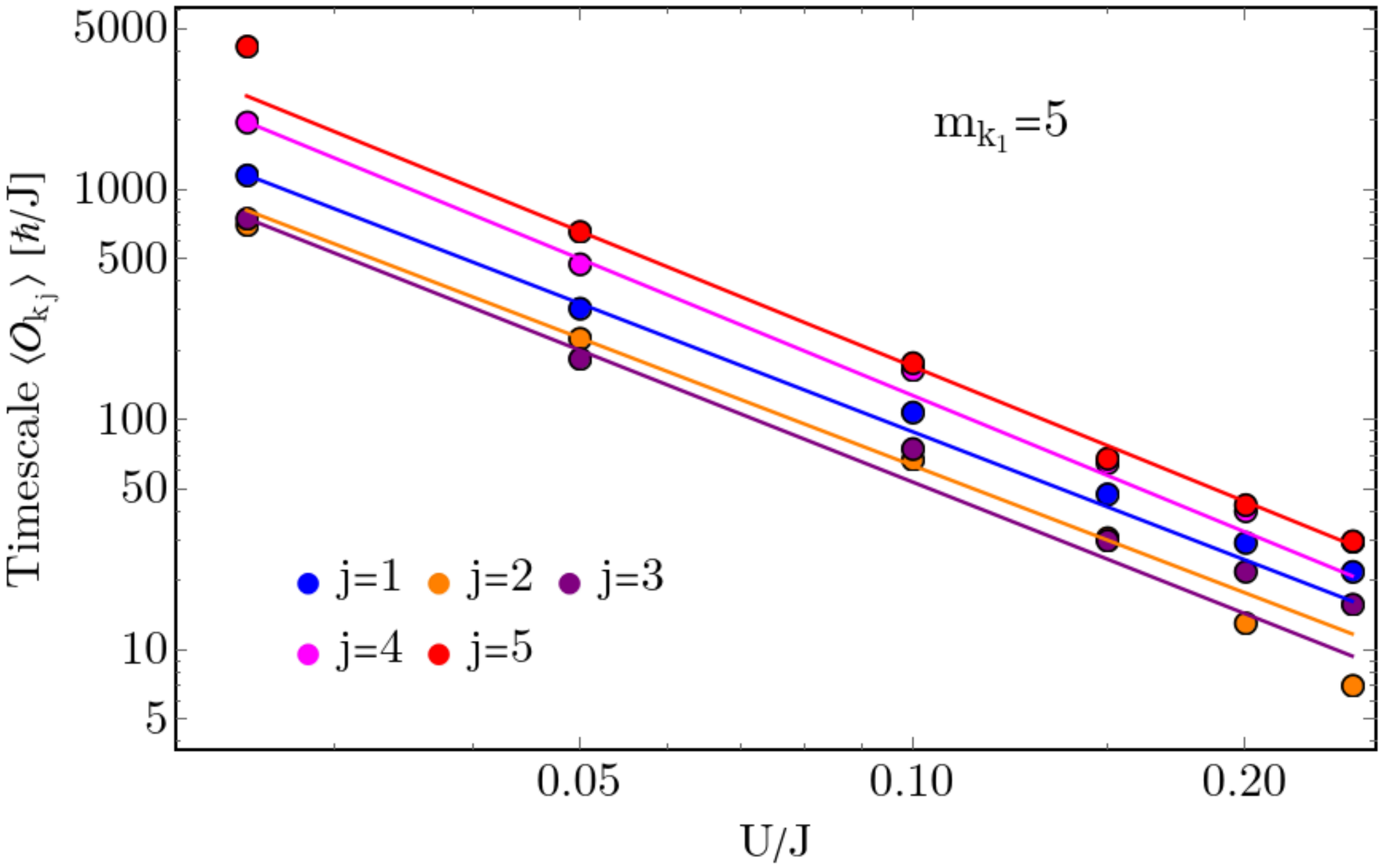}
\caption{The timescales obtained from the exponential fits of $\mathcal{O}_{k_j}$ as a function of $U$, for the data presented in Fig.~\ref{fig:breakingte1}. 
The lines represent a fit of the timescale dependence on the interaction with an algebraic decay $\propto U^{-\alpha}$, we obtain the following exponents: $j=1$, $\alpha=1.85 \pm 0.05$; $j=2$, $\alpha=1.84 \pm 0.07$; $j=3$, $\alpha=1.90 \pm 0.09$; $j=4$, $\alpha=1.97 \pm 0.06$; $j=1$, $\alpha=1.95 \pm 0.04$.
The parameters are chosen to be $L=10$, $N=5$, $\hbar\Omega\sqrt{N}/J=4.47$, $\hbar\delta/J=2$, and $\Gamma/J=15$.   
}
\label{fig:breakingtime}
\end{figure}

\begin{figure*}[hbtp]
\centering
\includegraphics[width=.98\textwidth]{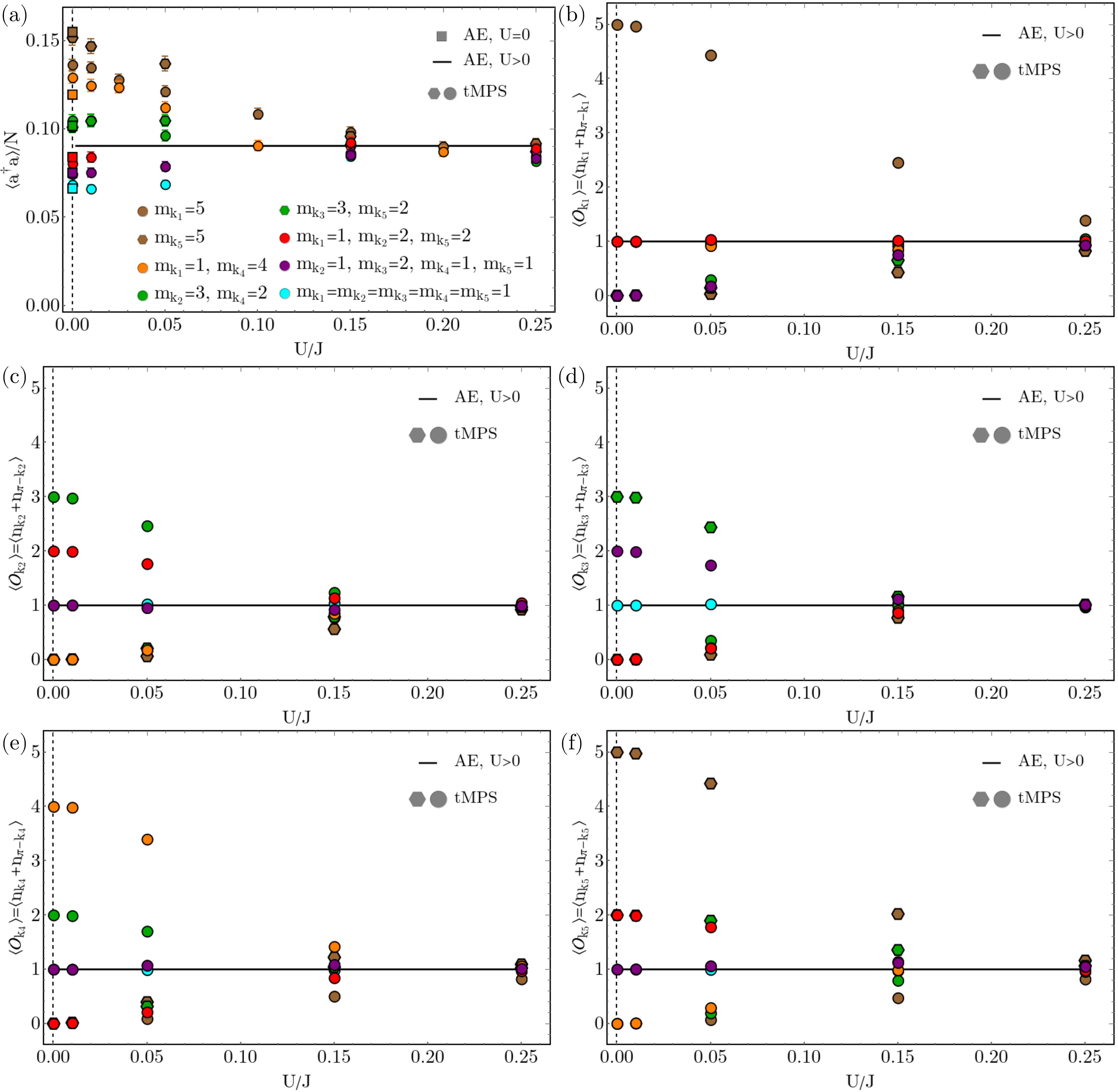}
\caption{The dependence on the interaction strength $U$ of (a) the scaled photon number, $\langle a^\dagger a\rangle/N$, and [(b)-(f)] the expectation value of $\mathcal{O}_{k_j}$, $j=1..5$ using tMPS at time $tJ=49.75\hbar$ and many-body adiabatic elimination (AE), for different symmetry sectors. The symbols identifying each symmetry sector are consistent in all panels. The parameters are chosen to be $L=10$, $N=5$, $\hbar\Omega\sqrt{N}/J=4.47$, $\hbar\delta/J=2$, and $\Gamma/J=15$.  
}
\label{fig:breaking}
\end{figure*}

In Figs.~(\ref{fig:breakingte1}-\ref{fig:breaking}) we present additional data complementing Fig.~3 comparing the many body adiabatic elimination results taking the kinetic term as the perturbation and the numerical exact tMPS results at large dissipation strengths.

In Fig.~\ref{fig:breakingte1} and Fig.~\ref{fig:breakingte2} the time evolution of the scaled photon number, $\langle a^\dagger a\rangle/N$, and of the conserved quantities, $\langle \mathcal{O}_{k_j}\rangle$, are shown for different values of the interaction strength for two additional symmetry sectors are plotted. The dissipation strength has been chosen large, $\Gamma/J=15$, such that we can compare to the results of the many body adiabatic elimination with the kinetic term  as the perturbation.
As for the symmetry sector presented in Fig.~3, we see that, at finite interaction, the late time behavior is nicely described by an exponential decay towards the many body adiabatic elimination state, $\rho_{\text{mix}}$ [see Eq.~(\ref{eq:ss_gen})]. We capture this by performing an exponential fit for $\langle \mathcal{O}_{k_j}\rangle -\langle \mathcal{O}_{k_j}\rangle _{\rho_{\text{mix}}}$, $\propto e^{-t/\tau}$. We see that the fit work very well in most cases which supports the decay towards the steady state $\rho_{\text{mix}}$. The decay time $\tau$ gives the timescale for reaching the steady state. The deviations seen in the curves for the strongest interaction are of the order of the statistical uncertainty of the Monte-Carlo sampling of the different trajectories. The timescales corresponding to Fig.~\ref{fig:breakingte1} are shown in Fig.~3(e), for the photon number, and in Fig.~\ref{fig:breakingtime}, for the conserved quantities, and support the decay of $\tau \propto 1/U^2$. We note that for the symmetry sector considered in Fig.~\ref{fig:breakingte1} we do not have enough data to extract the exponent.

In Fig.~\ref{fig:breaking} we look at the finite time value at time $tJ=49.75\hbar$ of the scaled photon number, $\langle a^\dagger a\rangle/N$, and all conserved quantities, $\langle \mathcal{O}_{k_j}\rangle $, as a function of the interaction strength for different symmetry sectors. Part of the data presented is overlapping with the data in Fig.~3 (a) and (b), but here additional symmetry sectors are shown and all conserved quantities $\langle \mathcal{O}_{k_j} \rangle$. 

In the presence of the strong symmetry, at $U=0$ (marked by a dashed vertical line in Fig.~\ref{fig:breaking}), we compare the tMPS results with the expectations value computed with the state $\rho_{\text{K,st}}$, Eq.~(\ref{eq:momss}), and we obtain very good agreement in all sectors. We note that for the state $\rho_{\text{K,st}}$  the expectation of the photon number depends only on the distribution of the particles in the single particle sectors and not the particular momentum values the particles have, i.e. the sectors $(m_{k_1}=5)$ and $(m_{k_5}=5)$ have the same photon number. At finite interaction we expect an agreement between the tMPS results and the state $\rho_{\text{mix}}$, Eq.~(\ref{eq:ss_gen}), as we can observe for $U/J\gtrsim 0.2$. The deviations at lower $U$ we attribute to the fact that the numerical results are taken at finite time and not in the steady state (see Figs.~\ref{fig:breakingte1}, \ref{fig:breakingte2}), since the exponential fit in Fig.~\ref{fig:breakingte1} and Fig.~\ref{fig:breakingte2} approaches the correct steady state value.

\setcounter{equation}{0}
\renewcommand{\theequation}{C.\arabic{equation}}

\section{\label{appC}Details of the \MakeLowercase{t}MPS method for the coupled
photon-atom system}

The numerically exact results are obtained with a matrix product state (MPS) method developed for the simulation of the time evolution of the dissipative master equation, Eqs.~(1)-(2) in the main text, for the cavity-atoms coupled systems \cite{HalatiKollath2020,HalatiKollath2020b}. The details regarding the implementation and benchmarking of the method are presented in Ref.~\cite{HalatiKollath2020b}. The method is based on the stochastic unravelling of the master equation with quantum trajectories \cite{DalibardMolmer1992, GardinerZoller1992, Daley2014} and a variant of the quasi-exact time-dependent matrix product state (tMPS) employing the Trotter-Suzuki decomposition of the time evolution propagator \cite{WhiteFeiguin2004, DaleyVidal2004, Schollwoeck2011} and the dynamical deformation of the MPS structure using swap gates \cite{StoudenmireWhite2010, Schollwoeck2011, WallRey2016}.

The convergence of our results is sufficient for at least 500 quantum trajectories in the Monte Carlo sampling, the truncation error goal of $10^{-12}$, the time-step of $dtJ = 0.0125 \hbar$ for the parameters used in Fig.~1 and $dtJ = 0.00625 \hbar$ for the other parameters considered in this work, and an adaptive cutoff of the local Hilbert space of the photon mode between
$N_\text{pho}= 20$  and $N_\text{pho}= 10$.

\setcounter{equation}{0}
\renewcommand{\theequation}{D.\arabic{equation}}
\setcounter{figure}{0}
\renewcommand{\thefigure}{D\arabic{figure}}

\section{\label{appD}Dissipative freezing}

\begin{figure*}[hbtp]
\centering
\includegraphics[width=\textwidth]{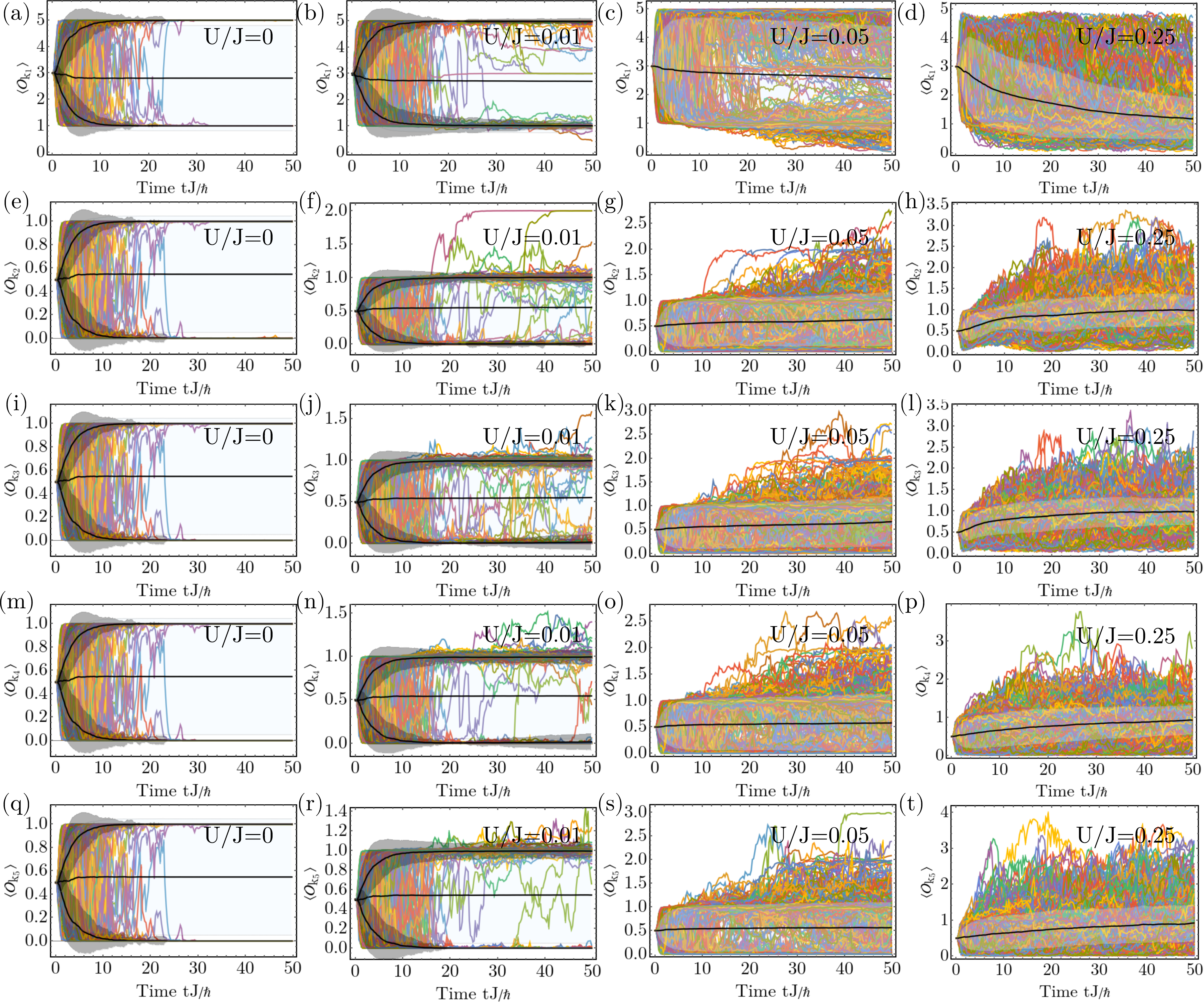}
\caption{Time evolution of $\mathcal{O}_k$ for the single quantum trajectories sampled in the Monte Carlo average for different interaction strengths $U$. The initial state consists in an equal superposition between states for the sectors $\left(m_{k_1}=5\right)$ and $\left(m_{k_1}=1,m_{k_2}=1,m_{k_3}=1,m_{k_4}=1,m_{k_5}=1\right)$. In each panel there are 1000 trajectories plotted, the black represent the Monte Carlo average, either for the full set of trajectories, or averaged separately depending on the final value, we shade the interval of one standard deviation away from the average, with light blue for the full average and light gray for the separate averages. 
The parameters used are $L=10$, $N=5$, $\hbar\delta/J=2$, $\hbar\Omega\sqrt{N}/J=4.47$, and $\hbar\Gamma/J=15$.
The standard deviation is defined as $\sigma(\mathcal{O}_k(t))=\sqrt{\frac{1}{R}\sum_{r=1}^R \left(\bra{\psi_r(t)}\mathcal{O}_k\ket{\psi_r(t)}-\langle\langle\mathcal{O}_k\rangle\rangle\right)^2}$, where $R$ is the total number of trajectories, $\ket{\psi_r(t)}$ the time-evolved wave function of the trajectory labeled by $r$ and $\langle\langle\mathcal{O}_k\rangle\rangle$ the statistical average over all trajectories.
}
\label{fig:breaking2}
\end{figure*}

In this appendix we describe how the phenomenon of dissipative freezing occurs for the special case that the system satisfies $\left[H,L^\dagger L\right]=0$.

The time-evolution of a single quantum trajectory is given by the following time-evolution operator
\begin{align} 
\label{eq:nonunitary_evolution}
U(t,t_0)=\frac{1}{\mathcal{M}}e^{-i \tilde{H}(t-t_N)/\hbar}\prod_{j=N}^{1} L~ e^{-i \tilde{H}(t_j-t_{j-1})/\hbar},
\end{align}
where $\mathcal{M}$ is the normalization constant, $L$ the jump operator, and $\{t_1,...,t_N\}$ are the stochastically sampled times when a quantum jump occurs. The effective non-Hermitian Hamiltonian is
\begin{align} 
\label{eq:effHam}
\tilde{H}=H-\frac{i}{2}\hbar\Gamma L^\dagger L.
\end{align} 

In order to analyze the phenomenon of dissipative freezing we can look at the evolution of one of the symmetry generators $\mathcal{O}_k$ in a single quantum trajectory,
\begin{align} 
\label{eq:sym_nonunitary_evolution}
\langle \mathcal{O}_k(t) \rangle_{traj} &=\bra{\psi_0} U^\dagger(t,t_0) \mathcal{O}_k U(t,t_0) \ket{\psi_0} \\
&=\bra{\psi_0} U^\dagger(t,t_0)U(t,t_0) \mathcal{O}_k  \ket{\psi_0}, \nonumber
\end{align}
with $\ket{\psi_0}$ the initial state. If $\ket{\psi_0}$ is within one symmetry sector and is an eigenstate of $\mathcal{O}_k$, the expectation value within the single trajectory $\langle \mathcal{O}_k(t) \rangle_{traj}$ will not evolve in time, as neither the jump operator, or the Hamiltonian can change the symmetry sector. In contrast, if the initial state is taken as a superposition with contributions from different symmetry sectors, then, in principle, both the jump operator or the evolution with the effective Hamiltonian can change the weights of these contributions. This implies that $\langle \mathcal{O}_k(t) \rangle_{traj}$ will evolve in time using a single quantum trajectory and only the Monte Carlo average will be constant.

In the case the systems satisfies $\left[H,L^\dagger L\right]=0$ one can get a better insight as 
\begin{align} 
\label{eq:sym_nonunitary_evolution2}
 U^\dagger(t,t_0)U(t,t_0)=\frac{1}{\mathcal{M}^2}e^{-\Gamma L^\dagger L(t-t_0)}(L^\dagger L)^n,
\end{align}
with $n$ the number of quantum jumps that occur up to time $t$. Here the evolution of $\mathcal{O}_k$ will only depend on the number of quantum jumps that occur up to time $t$.
This includes the particular case of $L^\dagger L=\mathbb{I}$ when due to the normalization in each jump $U^\dagger(t,t_0)U(t,t_0)=\mathbb{I}$ and $\langle \mathcal{O}(t) \rangle_{traj}$ is constant. The system considered in Ref.~\cite{MunozPorras2019} is also included in this situation, as the authors prove that dissipative freezing always occurs if $H\propto L\propto \mathcal{O}$. 

For the coupled atom-cavity system that we consider in this work, Eqs.~(1)-(2) in the main text, the condition $\left[H,L^\dagger L\right]\neq 0$ is not satisfied and the arguments given above are not directly applicable. Nevertheless, we show numerically that the dissipative freezing occurs even for this more involved case. 

In Fig.~\ref{fig:breaking2} we extend the data presented in Fig.~2, by plotting the expectation value of all the generators of the strong symmetry, $\langle\mathcal{O}_{k_j}\rangle$, $j=1..5$, in time for 1000 single trajectories. 
The initial state is an equal superposition of a state from the sector $(m_{k_1}=5)$ and the sector $(m_{k_1}=1, m_{k_2}=1, m_{k_3}=1, m_{k_4}=1, m_{k_5}=1)$. We can observe in the first column of Fig.~\ref{fig:breaking2} that the phenomenon of dissipative freezing can be identified in the evolution of each of the symmetry generators, as for times $tJ\gtrsim 40\hbar$, all trajectories evolved to one of the two symmetry sectors and the Monte Carlo average of the trajectories stays constant throughout the following time-evolution. If we slightly turn on the interaction and break the strong symmetry (see second column of Fig.~\ref{fig:breaking2} for $U/J=0.01$) we see that at short and intermediate time scales the behavior of the quantum trajectories is very similar to dissipative freezing. Thus we can infer that the approximate strong symmetry still affects the short-time dynamics. If we increase the interaction even further, $U/J\geq 0.05$, the mixing of the trajectories starts earlier and the dissipative freezing effects are washed out. 

\end{document}